\documentclass[a4paper,10pt,twoside]{cpc-hepnp-ma}

\usepackage{multicol}
\usepackage{graphicx}
\usepackage{booktabs}
\usepackage{amssymb,bm,mathrsfs,bbm,amscd}
\usepackage[tbtags]{amsmath}
\usepackage{lastpage}
\usepackage{CJK}

\begin{document}

\fancyhead[co]{\footnotesize ZHOU \& MA : New theory of Lorentz
violation from a general principle}

\footnotetext[0]{Received 9 March 2011, Revised 30 March 2011}

\title{New theory of Lorentz violation from a general principle}

\author{%
      ZHOU Lingli$^{1;1)}$\email{zhoull@pku.edu.cn}%
\quad MA Bo-Qiang$^{1;2)}$\email{mabq@pku.edu.cn}%
} \maketitle

\address{%
$^1$ School of Physics and
State Key Laboratory of Nuclear Physics and Technology, Peking
University, Beijing 100871, China
}

\begin{abstract}
We report that a general principle of physical independence of
mathematical background manifolds brings a replacement of common
derivative operators by co-derivative ones. Then we obtain a new
Lagrangian for the ordinary minimal standard model with supplementary
terms containing the Lorentz invariance violation information
measured by a new matrix, denoted as the Lorentz invariance violation
matrix. We thus provide a new fundamental theory to study Lorentz
invariance violation effects consistently and systematically.
\end{abstract}



\begin{multicols}{2}

\section{Introduction}

Lorentz symmetry is one of the most significant and fundamental
principles in physics, and it contains two aspects, namely Lorentz
covariance and Lorentz invariance. Currently, there is
an increasing interest in Lorentz invariance Violation (LV or LIV)
both theoretically and experimentally~\cite{ShaoMa10}. We report
here a fundamental theory to describe the LV effects based on a
basic principle. Similar to the principle of relativity, which
requires that the equations describing the laws of physics have the
same form in all admissible frames of reference, we propose a
physical independence principle that the equations describing the
laws of physics have the same form in all admissible mathematical
manifolds. We show that such a principle leads to the following
replacement of the ordinary partial $\partial_{\alpha}$ and the
covariant derivative $D_{\alpha}$
\begin{equation}\label{eqn:substitution}
\partial^{\alpha} \rightarrow M^{\alpha\beta}\partial_{\beta},\quad
D^{\alpha}\rightarrow M^{\alpha\beta}D_{\beta},
\end{equation}
where $M^{\alpha\beta}$ is a local matrix. We first introduce the
general principle, and then show that such principle leads to new
supplementary terms violating Lorentz invariance in the Standard Model.

\section{Principle of independence}
Principle: Under any one-to-one transformation $X\rightarrow
X'=f(X)$ of background mathematical manifolds, the corresponding
transformation $\varphi(\cdot)\rightarrow \varphi'(\cdot)$ of an
arbitrary physical field $\varphi(X)$ should satisfy
\begin{equation}\label{eqn:PFIMBM}
\varphi'(X')=\varphi(X).
\end{equation}
This statement actually makes the field $\varphi(X)$ represent a
physical distribution rather than a mathematical function. A unique
reality can be described in many ways
$\varphi(X),\varphi'(X'),\varphi''(X''),\ldots$ mathematically, but
the physical essence remains unchanged, saying independence or
invariance of mathematical descriptions. So (\ref{eqn:PFIMBM}) just
claims Physical Independence or Physical Invariance (PI) of
mathematical background manifolds. What we do here is just to put a
physical requirement on a mathematical expression $\varphi(X)$.

For a given field $\varphi(X)$ satisfying (\ref{eqn:PFIMBM}), its
derivative field is ordinarily defined as
\begin{equation}\label{def1}
\pi(X)=\partial_{X}\varphi(X).
\end{equation}
When $\pi(X)$ is a physical field, we should require the condition
$\pi'(X')=\pi(X)$ according to PI. Therefore we need to check
whether the definition (\ref{def1}) of the derivative field
satisfies (\ref{eqn:PFIMBM}). One easily finds
\begin{eqnarray}\label{eqn:momentum}
\pi(X)&=&\partial_{X}\varphi(X)=\partial_{X}\varphi'(X') \nonumber \\
      &=&\partial_{X}f(X)\ast\partial_{X'}\varphi'(X')\nonumber\\
      &=&F(\partial_{X'})\varphi'(X'),\nonumber
\end{eqnarray}
in which $F(\cdot)=\partial_{X}f(X)\ast\cdot$, where $F(\cdot)$ is
linear to ``$\cdot$". From the definition (\ref{def1}),
$\pi'(X')=~\partial_{X'}\varphi'(X')$, we see that
$\pi'(X')\neq\pi(X)$. So this definition of the derivative field
$\pi(X)$ of a physical field $\varphi(X)$ does not satisfy PI. The
reason is due to the derivative with respect to the manifold $X$. So
we define the derivative field as
\begin{equation}\label{def2}
\pi(X)=M(\partial_{X})\varphi(X),
\end{equation}
which indicates that $M(\cdot)$ is local and has the 
transformation property
\begin{equation}
M(\cdot)\rightarrow M'(\cdot)=M(F(\cdot)),
\end{equation}
thus we have
\begin{eqnarray}
\pi(X)&=&M(\partial_{X})\varphi(X)\nonumber\\
      &=&M(\partial_{X}f(X)\ast\partial_{X'})\varphi'(X')\nonumber\\
      &=&M(F(\partial_{X'}))\varphi'(X')\nonumber\\
      &=&M'(\partial_{X'})\varphi'(X')\nonumber\\
      &=&\pi'(X').\nonumber
\end{eqnarray}
According to (\ref{eqn:PFIMBM}), $\pi(X)$ is indeed a physical field
with the new definition (\ref{def2}). The covariant derivative
$D_{X}$ has the same problem as $\partial_{X}$ and can be handled in
a similar manner. Thus we obtain our replacement for the ordinary
$\partial_{X}$ and the covariant derivative $D_{X}$ by
\begin{displaymath}
\partial_{X}\rightarrow M(\partial_{X}),\quad
D_{X}\rightarrow M(D_{X}),
\end{displaymath}
whose explicit matrix form  is
\begin{displaymath}
\partial^{K}\rightarrow M^{KJ}\partial_{J},\quad
D^{K}\rightarrow M^{KJ}D_{J}.
\end{displaymath}
We need to point out that the above derivation is handled in
Geometric Algebra $ \mathcal{G}$ (or Clifford Algebra) and Geometric
Calculus (see, e.g., Refs. \cite{GAtoGC,Doran03}), where the general
element is called a multivector. Addition and various products of
two multivectors are still a multivector, i.e., Geometric Algebra is
closed. Different variables in physics, such as scalar, vector,
tensor, spinor, twistor, matrix, etc., can be described by the
corresponding types of multivectors in a unified form in Geometric
Algebra (see Ref.~\cite{Ma10} for a more detailed argument). So
$\varphi(X) \in \mathcal{G}$  is a multivector-valued function of a
multivector variable $X \in \mathcal{G}$. In the following
discussions, when we consider the space-time, which is part of the
general Geometric Algebra space, $x$ is adopted instead of $X$, and
the indices are denoted by $\alpha,\beta$ instead of $K,J$.

The result (\ref{eqn:substitution}) is similar to the gauge idea by
Yang and Mills \cite{Yang54} from a basic consideration. When a local
symmetry is considered, one has to replace a common partial
$\partial_{\alpha}$ with a covariant derivative $D_{\alpha}$ to
retain the invariance of a Lagrangian under the local gauge
transformation. Requiring the property (\ref{eqn:PFIMBM}) for an
arbitrary field, we must introduce the local matrix
$M^{\alpha\beta}$ to make the common $\partial^{\alpha}$ a physical
co-derivative operator $M^{\alpha\beta}\partial_{\beta}$. The
combination of the above two general considerations brings about the
new covariant co-derivative operator $D^{\alpha}\rightarrow
M^{\alpha\beta}D_{\beta}$, which is essential for the natural
introduction of LV terms in the Standard Model.

\section{Standard model supplement}
With the above considerations, we focus on the physical implications
and consequences from the new introduced co-derivative
$M^{\alpha\beta}\partial_{\beta}$ and covariant co-derivative
$M^{\alpha\beta}D_{\beta}$. The effective Lagrangian
$\mathcal{L}_{\mathrm{SM}}$ of the minimal standard model is
composed of four parts
\begin{eqnarray}
\mathcal{L}_{\mathrm{SM}}&=& \mathcal{L}_{\mathrm{G}}+\mathcal{L}_{\mathrm{F}}+\mathcal{L}_{\mathrm{H}}+\mathcal{L}_{\mathrm{HF}},\label{eqn:SM}\\
\mathcal{L}_{\mathrm{G}} &=& -\frac{1}{4}F^{a\alpha\beta } F_{\alpha\beta}^{a}, \label{eqn:SMG}\\
\mathcal{L}_{\mathrm{F}} &=&i\bar{\psi}\gamma^{\alpha}D_{\alpha}\psi, \label{eqn:SMF}\\
\mathcal{L}_{\mathrm{H}}&=& (D^{\alpha}\phi)^{\dag}D_{\alpha}\phi +
V(\phi),\label{eqn:SMHG}
\end{eqnarray}
where 
we omit the chiral differences, the summation of chirality and gauge
scripts.
 $\psi$ is the fermion field, $\phi$ is the Higgs
field, and $V(\phi)$ is the Higgs self-interaction.
$F_{\alpha\beta}^{a}=\partial_{\alpha}A_{\beta}^{a}-\partial_{\beta}A_{\alpha}^{a}
-gf^{abc}A_{\alpha}^{b}A_{\beta}^{c}$,
$D_{\alpha}=\partial_{\alpha}+igA_{\alpha}$ and
$A_{\alpha}=A_{\alpha}^{a}t^{a}$, with $A_{\alpha}^{a}$ being the
gauge field. $g$ is the coupling constant, and $f^{abc}$ and $t^{a}$
are the structure constants and generators of the corresponding
gauge group respectively. $\mathcal{L}_{\mathrm{HF}}$ is the Yukawa
coupling between the fermions and the Higgs field, and is not
related to derivatives, thus it remains unchanged under the
replacement (\ref{eqn:substitution}).

Under (\ref{eqn:substitution}) and a decomposition
$M^{\alpha\beta}=g^{\alpha\beta}+\Delta^{\alpha\beta}$ which will be
discussed later, the Lagrangians in (\ref{eqn:SMG})-(\ref{eqn:SMHG})
become
\begin{eqnarray}
\mathcal{L}_{\mathrm{G}} &=& -\frac{1}{4}
 (M^{\alpha\mu}\partial_{\mu}A^{a\beta}-M^{\beta\mu}\partial_{\mu}A^{a\alpha}
  -gf^{abc}A^{b\alpha}A^{c\beta})       \nonumber\\
  &\times&
 (M_{\alpha\mu}\partial^{\mu}A_{\beta}^{a}-M_{\beta\mu}\partial^{\mu}A_{\alpha}^{a}
  -gf^{abc}A_{\alpha}^{b}A_{\beta}^{c}) \nonumber\\
  &=&-\frac{1}{4}F^{a\alpha\beta }F_{\alpha\beta}^{a}+ \mathcal{L}_{\mathrm{GV}},\label{eqn:SMSG}\\
\mathcal{L}_{\mathrm{F}}
&=&i\bar{\psi}\gamma_{\alpha}M^{\alpha\beta}D_{\beta}\psi
  =i\bar{\psi}\gamma^{\alpha}D_{\alpha}\psi+  \mathcal{L}_{\mathrm{FV}}, \label{eqn:SMSF} \\
\mathcal{L}_{\mathrm{H}}&=&(M^{\alpha\mu}D_{\mu}\phi)^{\dag}M_{\alpha\nu}D^{\nu}\phi
  + V(\phi)                                                                \nonumber \\
  &=&(D^{\alpha}\phi)^{\dag}D_{\alpha}\phi + V(\phi)+ \mathcal{L}_{\mathrm{HV}},\label{eqn:SMSHG}
\end{eqnarray}
with $M^{\alpha\beta}$ being the real matrix to maintain the Lagrangian
hermitian. The last three terms $\mathcal{L}_{\mathrm{GV}}$,
$\mathcal{L}_{\mathrm{FV}}$ and $\mathcal{L}_{\mathrm{HV}}$ of the
equations mentioned above are the supplementary terms for the ordinary Standard
Model. The explicit forms of these terms are
\begin{eqnarray}
\mathcal{L}_{\mathrm{GV}}&=&
-\frac{1}{2}\Delta^{\alpha\beta}\Delta^{\mu\nu}(g_{\alpha\mu}\partial_{\beta}
A^{a\rho}\partial_{\nu}A_{\rho}^{a}-\partial_{\beta}A_{\mu}^{a}
\partial_{\nu}A_{\alpha}^{a})    \nonumber \\
   & & -F_{\mu\nu}^{a}\Delta^{\mu\alpha}\partial_{\alpha}A^{a\nu},\label{eqn:GV}  \\
\mathcal{L}_{\mathrm{FV}}&=&
 i\Delta^{\alpha\beta}\bar{\psi}\gamma_{\alpha}\partial_{\beta}\psi
  -g\Delta^{\alpha\beta}\bar{\psi}\gamma_{\alpha}A_{\beta}\psi, \label{eqn:FV} \\
\mathcal{L}_{\mathrm{HV}}&=&
(g_{\alpha\mu}\Delta^{\alpha\beta}\Delta^{\mu\nu}+\Delta^{\beta\nu}+\Delta^{\nu\beta})
  (D_{\beta}\phi)^{\dag}D_{\nu}\phi.  \label{eqn:HGV}
\end{eqnarray}
Thus we obtain a new effective Lagrangian for the Standard Model with
new supplementary terms (SMS), denoted by $\mathcal{L}_{\mathrm{SMS}}$
\begin{eqnarray}
&\mathcal{L}_{\mathrm{SMS}}&= \mathcal{L}_{\mathrm{SM}} + \mathcal{L}_{\mathrm{LV}}, \label{eqn:SMS}\\
&\mathcal{L}_{\mathrm{LV}}&=
\mathcal{L}_{\mathrm{GV}}+\mathcal{L}_{\mathrm{FV}}+\mathcal{L}_{\mathrm{HV}},\label{eqn:LIV}
\end{eqnarray}
where $\mathcal{L}_{\mathrm{SMS}}$ satisfies the Lorentz covariance
(SO$^{+}$(1,3)), the gauge symmetry invariance of
SU(3)$\times$SU(2)$\times$U(1) and invariance under the requirement
of PI (\ref{eqn:PFIMBM}), under which $\mathcal{L}_{\mathrm{SM}}$
cannot remain unchanged in a general situation.

We can have a better view of 
SMS here. The elements of $M^{\alpha\beta}$ are mass dimensionless,
natural for the sign of testifying the Lorentz invariance, and they are
not global constants generally. All of the LV terms are expressed in
$\mathcal{L}_{\mathrm{LV}}$, and the LV
information~\cite{Liberati09} is measured by a single concise matrix
$\Delta^{\alpha\beta}$, which is convenient for a systematic study
of the LV effects.  To determine whether the Lorentz invariance holds
exactly (see (\ref{LIMmeaning})), further work is needed to analyze
the effective Lagrangian (\ref{eqn:SMS}) of QED, QCD and EW
(ElectroWeak) fields, and more experiments are needed to determine
the magnitude of the elements in the matrix $M^{\alpha\beta}$.

\section{Lorentz violation matrix}
Now, let us focus on the new local matrix $M^{\alpha\beta}$, of
which the vacuum expectation value is used for the coupling
constants in (\ref{eqn:GV}), (\ref{eqn:FV}) and  (\ref{eqn:HGV}). We
divide $M^{\alpha\beta}$ into two parts
\begin{equation}\label{eqn:Mdivide}
M^{\alpha\beta}=g^{\alpha\beta}+\Delta^{\alpha\beta},
\end{equation}
with $g^{\alpha\beta}$ being the space-time metric. The remaining
$\Delta^{\alpha\beta}$  contains the information to judge whether
the Lorentz invariance is kept or not
\begin{equation}\label{LIMmeaning}
\Delta^{\alpha\beta}= \left\{
  \begin{array}{ll}
            0           & \textrm{no LV}, \\
    \rightarrow0        & \textrm{small LV}, \\
    \textrm{otherwise}  & \textrm{large LV}. \\
  \end{array}
\right.
\end{equation}
So $\Delta^{\alpha\beta}$ represents to what degree the Lorentz
invariance is exact, and we call it the Lorentz invariance Violation
Matrix (LVM). Intuitively, the smaller the elements of
$\Delta^{\alpha\beta}$ are, the better the physics law holds Lorentz
invariant. In this way, the LVM is similar to the CKM matrix which
signals generation mixing and CP violation~\cite{Cabibo63,KM73}, and
it signals LV.

Generally, $\Delta^{\alpha\beta}$ might be particle-type dependent.
If we use the vacuum expectation value of $\Delta^{\alpha\beta}$ for
the coupling constants in the corresponding effective Lagrangian,
not all of the 16 degrees of freedom of $M^{\alpha\beta}$ are
physical. For the derivative field $M(\partial_{x})\varphi(x)$
(\ref{def2}) of an arbitrary given field, $\varphi(x)$ can be
rescaled to absorb one of the 16 degrees of freedom so that only 15
are left. When more fields are involved, there is only one
degree of freedom that can be reduced from a rescaling consideration
for all fields. Thus for generality, we may keep all 16 degrees of
freedom in $M^{\alpha\beta}$ for a specific particle in our study.

With the rapid development of laboratory
experiments~\cite{Muller07} and astronomical
observations~\cite{Amelino98,Ellis00,Mattingly05,Amelino09,Ellis09,Xiao09,Shao10},
there will be more and more ways to determine the LVM
$\Delta^{\alpha\beta}$ phenomenologically. For example, we can get
the dynamical equations of fields such as modified Maxwell
equations, the Klein-Gordon equation and the Dirac equation as well as
various dispersion relations from the effective Lagrangian. As a
preliminary test for our construction, we consider the Dirac
equation for the fermion $\psi(x)$ through the substitution
(\ref{eqn:substitution}) or the Lagrangian (\ref{eqn:SMSF})
\begin{equation}\label{eqn:Dirac}
(i\gamma_{\alpha}M^{\alpha\beta}\partial_{\beta}-m)\psi(x)=0.
\end{equation}
Multiplying both sides by
$(i\gamma_{\alpha}M^{\alpha\beta}\partial_{\beta}+m)$, we obtain
\begin{equation}
(g_{\alpha\mu}M^{\alpha\beta}M^{\mu\nu}\partial_{\beta}\partial_{\nu}+m^2)\psi(x)=0,
\end{equation}
which is also a Klein-Gordon equation. So the dispersion relation is
\begin{equation}\label{eqn:disper}
p^2+g_{\alpha\mu}\Delta^{\alpha\beta}\Delta^{\mu\nu}p_{\beta}p_{\nu}+
2\Delta^{\alpha\beta}p_{\alpha}p_{\beta}=m^2,
\end{equation}
with the Fourier transformation $\psi(x)=\int \psi(p)e^{-ip\cdot
x}dp$ and (\ref{eqn:Mdivide}). The last two items of the left side
of (\ref{eqn:disper}) contain LV information. It is an extension
of the ordinary mass-energy relation $p^2=m^2$. Systematic analysis
of LV here can be offered by adopting the general expression for
$\Delta^{\alpha\beta}$. A special case is
\begin{equation}\label{eqn:LIMsimple}
\Delta^{\alpha\beta}=\mathrm{diag}(0,\xi,\xi,\xi),
\end{equation}
where we consider first only the diagonals for simplicity.  Then
(\ref{eqn:disper}) and (\ref{eqn:LIMsimple}) give
\begin{eqnarray}
E^2&=&(1-\delta)\vec{p}^2+m^2, \\
\delta&=&-\xi^2+2\xi. \nonumber
\end{eqnarray}
The photopion production of the nucleon in the GZK cutoff
observations gives an available energy threshold
$E\approx{10^{19}}$eV \cite{Bietenholz08}. For the proton, it can be
used here to determine the magnitude of the upper limit of $\xi$,
which is $10^{-23}$  from a rough estimate~\cite{Xiao09I}. The
details of analysis, which are presented in \cite{Ma10}, are omitted
here.

From the Lagrangian for free photons
\begin{eqnarray}
\mathcal{L}_{\mathrm{G}}&=&-\frac{1}{4}F^{\alpha\beta
}F_{\alpha\beta}
-F_{\mu\nu}\Delta^{\mu\alpha}\partial_{\alpha}A^{\nu}\label{Lagrangian} \nonumber\\
&-&\frac{1}{2}\Delta^{\alpha\beta}\Delta^{\mu\nu}(g_{\alpha\mu}\partial_{\beta}
A^{\rho}\partial_{\nu}A_{\rho}-\partial_{\beta}A_{\mu}
\partial_{\nu}A_{\alpha}),
\end{eqnarray}
we get the modified Maxwell equation (or motion equation)
\begin{equation}\label{maxwell1}
\Pi^{\gamma\rho}A_{\rho}=0,
\end{equation}
where $\Pi^{\gamma\rho}$ is also the inverse of the photon
propagator
\begin{eqnarray}
\Pi^{\gamma\rho}
&=&-g^{\gamma\rho}\partial^2+\partial^{\gamma}\partial^{\rho}\nonumber\\
&+&\Delta^{\gamma\alpha}\partial^{\rho}\partial_{\alpha}+\Delta^{\rho\alpha}\partial^{\gamma}\partial_{\alpha}
+\Delta^{\gamma\beta}\Delta^{\rho\nu}\partial_{\beta}\partial_{\nu}\nonumber\\
&-&g^{\gamma\rho}(2\Delta^{\mu\alpha}\partial_{\mu}\partial_{\alpha}
+g_{\alpha\mu}\Delta^{\alpha\beta}\Delta^{\mu\nu}\partial_{\beta}\partial_{\nu}).
\end{eqnarray}
The term $\partial^{\gamma}\partial^{\rho}$ is symmetric for the indices $\gamma$ and $\rho$. So are
$\Delta^{\gamma\alpha}\partial^{\rho}\partial_{\alpha}$ and
$\Delta^{\rho\alpha}\partial^{\gamma}\partial_{\alpha}$. Hence the above three terms can be omitted
under the constraint of the Lorentz gauge condition
$\partial^{\alpha}A_{\alpha}=0$ for the gauge field.
With the Fourier decomposition $A_{\rho}=\int dp
A_{\rho}(p)e^{-ip\cdot x}$ and the Lorentz gauge condition
we can re-write
Eq.~(\ref{maxwell1}) as $$\Pi^{\gamma\rho}(p)A_{\rho}(p)=0,$$ where
\begin{eqnarray}
\Pi^{\gamma\rho}(p)&=&g^{\gamma\rho}(p^2+g_{\alpha\mu}\Delta^{\alpha\beta}\Delta^{\mu\nu}p_{\beta}p_{\nu}+2\Delta^{\alpha\beta}p_{\alpha}p_{\beta})\nonumber\\
&-&\Delta^{\gamma\beta}\Delta^{\rho\nu}p_{\beta}p_{\nu},\nonumber
\end{eqnarray}
which is the inverse of the free photon propagator in the momentum
space. A general parametrization for $p_{\alpha}$ can be done with
spherical coordinates, so $p_{\alpha}$ can be expressed as $(E$,
$-|\vec{p}|\sin{\theta}\cos{\phi}$,
$-|\vec{p}|\sin{\theta}\sin{\phi}$, $-|\vec{p}|\cos{\theta})$, where
the light speed constant $c=1$ is adopted. We find that there is a
zero eigenvalue and a corresponding eigenvector $A_{\rho}(p)$ for
the matrix $\Pi^{\gamma\rho}(p)$. So the determinant must be zero
for the existence of the solution $A_{\rho}(p)$
\begin{equation}\label{det}
\mathrm{det}(\Pi^{\gamma\rho}(p))=0.
\end{equation}
Then we have the equation
$$\sum_{i=0}^{8}\lambda_{i}(\Delta^{\alpha\beta},\theta,\phi)E^{i}|\vec{p}|^{8-i}=0.$$
The coefficient $\lambda_{i}(\Delta^{\alpha\beta},\theta,\phi)$ is a
variable with respect to the LVM $\Delta^{\alpha\beta}$ and the
angles $\theta$ and $\phi$. So there are 8 real solutions  for
$E(|\vec{p}|)$ at most, and in general there are no analytical
solutions to a general high order linear equation. But for some
simple cases of the LVM $\Delta^{\alpha\beta}$, we expect some
analytical solutions for $E$. Anyway, $E$  can be solved formally as
$E=f_{i}(\Delta^{\alpha\beta},\theta,\phi)|\vec{p}|$, for $i=1\ldots
N$, and $1\leq N \leq 8$. $f_{i}(\Delta^{\alpha\beta},\theta,\phi)$
is a real variable and is independent of the momentum magnitude
$|\vec{p}|$ because the photon is massless in the Lagrangian of
Eq.~(\ref{Lagrangian}). So the free photon velocity is
\begin{equation}\label{groupSpeed}
c_{\gamma
i}\equiv\frac{dE}{d|\vec{p}|}=f_{i}(\Delta^{\alpha\beta},\theta,\phi),\quad
\textrm{for } i=1\ldots N,\quad 1\leq N \leq 8,
\end{equation}
which means: i) The free photon propagates in the space with at most
8 group velocities; ii) For each mode, the light speed $c_{\gamma
i}$ might be azimuth dependent and not a constant. As we know,
the light spreads with different group velocities for different
directions in the anisotropic media in optics. By analogy, we may
view the space-time as a kind of media intuitively. However, there
are essential differences between the optical case and the photon
case here, because all the consequences of the $N$ modes and the
light speed anisotropy are the result of the Lorentz invariance
violation or the space-time anisotropy suggested by the new theory.

Phenomenologically, the one-way
experiment~\cite{Gurzadyan05,Gurzadyan07,Gurzadyan10,Bocquet10}
performed at the GRAAL facility of the European Synchrotron
Radiation Facility (ESRF) in Grenoble, reported results on the light
speed anisotropy by Compton scattering of high-energy electrons on
laser photons. A detailed analysis, which will be given
elsewhere~\cite{Ma10b}, shows that the azimuthal distribution of the
GRAAL data can be elegantly explained by our new theory.

\section{Conclusion}
In summary, with a general requirement of physical independence or
physical invariance of mathematical background manifolds, we
introduce a replacement of common derivative operators by
co-derivative ones. This naturally brings about Lorentz invariance
violation terms in the Standard Model, and we get a Lorentz
invariance violation matrix $\Delta^{\alpha\beta}$, which can
describe the Lorentz invariance violation effects in a systematic
and consistent manner. The novel Lorentz invariance violation matrix
in this article has the following merits: i)~it is natural, because we
introduce it under a general consideration which makes a field
indeed a physical field without adding Lorentz invariance violation
terms by hand; ii)~it is systematic, since the information of Lorentz
invariance violation can be extracted from it uniquely; iii)~it is simple,
since the Lagrangian $\mathcal{L}_{\mathrm{SMS}}$ provides a
fundamental framework for elegant applications to experimental
problems. We thus provide a new fundamental theory to study the Lorentz
invariance violation effects consistently and systematically.

\acknowledgments{ This work was supported by the National Natural
Science Foundation of China (Grants No.~11021092, No.~10975003, and
No.~11035003) and by the Key Grant Project of Chinese Ministry of
Education (No. 305001). }

\end{multicols}

\vspace{-1mm}
\centerline{\rule{80mm}{0.1pt}}
\vspace{2mm}

\begin{multicols}{2}
\bibliographystyle{unsrt}

\end{multicols}
\end{document}